\documentclass[seceq]{ptptex}

\usepackage{graphicx}
\usepackage{cite}

\title{%        %You can use \\ for explicit line-break.
Constraining Galileon gravity from observational data with growth rate
}

\author{%       %Use \scshape for the family name.
Koichi \textsc{Hirano}$^{1,2,}$\footnote{E-mail: k\_hirano@tsuru.ac.jp}
, Zen \textsc{Komiya}$^{3}$
and Hisato \textsc{Shirai}$^{2}$
}

\inst{%     %Affiliation, neglected when [addenda] or [errata].
$^1$Department of Elementary Education, Tsuru University, Tsuru 402-8555, Japan \\ 
$^2$Department of Physics, Ichinoseki National College of Technology, Ichinoseki 021-8511, Japan \\
$^3$Department of Physics, Tokyo University of Science, Tokyo 162-8601, Japan
}

\abst{%         %This abstract is neglected when [addenda] or [errata].
We studied the cosmological constraints on the Galileon gravity obtained from observational data of the growth rate of matter density perturbations, the supernovae Ia (SN Ia), the cosmic microwave background (CMB), and baryon acoustic oscillations (BAO). For the same value of the energy density parameter of matter $\Omega_{m,0}$, the growth rate $f$ in Galileon models is enhanced, relative to the $\Lambda$CDM case, because of an increase in Newton's constant. The smaller $\Omega_{m,0}$ is, the more growth rate is suppressed. Therefore, the best fit value of $\Omega_{m,0}$ in the Galileon model, based only the growth rate data, is quite small. This is incompatible with the value of $\Omega_{m,0}$ obtained from the combination of SN Ia, CMB, and BAO data. On the other hand, in the $\Lambda$CDM model, the values of $\Omega_{m,0}$ obtained from different observational data sets are consistent. In the analysis of this paper, we found that the Galileon model is less compatible with observations than the $\Lambda$CDM model. This result seems to be qualitatively the same in most of the generalized Galileon models in which Newton's constant is enhanced.
}

\PTPindex{401, 453}  %Input the subject index(es) of your paper, 
\begin{document}

\maketitle

\section{INTRODUCTION}

The modifications to gravity at cosmological distances have recently received much attention as a possible explanation for cosmic acceleration. Many modified-gravity approaches have been proposed, including $f(R)$ gravity \cite{fel2010}, scalar-tensor theories \cite{hir2010c,hir2008a,har2008}, Gauss-Bonnet gravity \cite{car2005,noj2005}, the Dvali--Gabadazde--Porrati (DGP) braneworld model \cite{dva2000,hir2010a,hir2011a}, and Galileon gravity \cite{nic2009,def2009a,def2009b,cho2009,sil2009,kob2010a,kob2010b,gan2010,fel2010d,fel2010e,fel2010a,fel2010b,def2010,nes2010,kim2010}.

Modified-gravity models need to be constructed so that Newton gravity is recovered at short distances to ensure consistency with solar system experiments; the deviation from General Relativity (GR) can be allowed at large distances.

There are two ways to recover General Relativistic behavior in the high density regimes relevant to solar system experiments. The first is the so-called chameleon mechanism \cite{kho2004} in which the field mass is different, depending on the matter density in the surrounding environment. If the field is sufficiently heavy in the regions of high density, a spherically symmetrical body can have a thin-shell around its surface so that the effective coupling between the field and matter is suppressed outside the body. This mechanism can be at work in viable $f(R)$ gravity and scalar-tensor theory.

The second method of recovering GR in regions of high density is to introduce non-linear self-interactions in the form of a scalar field $\phi$. The simplest term of the self-interaction is $(\nabla\phi)^2\Box\phi$. There have been attempts to restrict the form of the Lagrangian by imposing "Galilean" symmetry, $\partial_\mu\phi\rightarrow\partial_\mu\phi+b_\mu$ \cite{nic2009}. The self-interaction of the form $(\nabla\phi)^2\Box\phi$ respects Galilean symmetry in the Minkowski background. The self-interaction term induces decoupling of the field $\phi$ from gravity at small scales by the so-called Vainshtein mechanism \cite{vai1972}. This allows the theory to recover general relativity for length scales smaller than the so-called Vainshtein radius \cite{vai1972}, which is consistent with solar system experiments (see Ref. \citen{kimura2012} for the implementation of the Vainshtein mechanism in these models). This term appears in the 4-dimensional effective theory of the DGP model. Unfortunately, the DGP model is plagued by the ghost problem \cite{koy2007} and is incompatible with cosmological observations \cite{xia2009}.

The presence of a nonlinear self-interaction of the form $\xi(\phi)(\nabla\phi)^2\Box\phi$, where $\xi(\phi)$ is a function of a scalar field $\phi$, as a part of Galileon gravity models has recently been studied \cite{nic2009,def2009a,def2009b,cho2009,sil2009,kob2010a,kob2010b,gan2010,fel2010d,fel2010e,fel2010a,fel2010b,def2010,nes2010,kim2010}. If $\xi(\phi)$ is constant, then this term respects Galilean symmetry in the Minkowski space-time, which keeps the equation of motion a second-order differential equation. This prevents the theory from introducing a new degree of freedom; perturbation of the theory does not lead to ghost or instability problems. Depending on the form of the function, some models do not have this "Galilean" symmetry. However, it is still possible to retain the desired properties. The equation of motion for the scalar field can remain of second order, which prevents the theory from having additional degrees of freedom.

Galileon gravity can induce self-accelerated expansion of the late-time Universe. Therefore, inflation models inspired by Galileon theory have been proposed \cite{kob2010c,miz2010,kam2010,fel2011,kob2011}. The evolution of matter density perturbations and conditions for the avoidance of ghosts and instabilities in Galileon models have also been studied \cite{sil2009,kob2010a,kob2010b,fel2010d,fel2010e,kim2010,fel2010c}. In Ref. \citen{kimura2011}, it is demonstrated that the integrated Sachs--Wolfe effect gives a stringent constraint on a subclass of the galileon model.

In this paper, we study the cosmological constraints using observational data of the growth rate of matter density perturbations, as well as supernovae Ia (SN Ia) \cite{ama2010}, the cosmic microwave background (CMB) \cite{kom2010}, and baryon acoustic oscillations (BAO) \cite{per2010}, for the action (\ref{action}) given below. In low-energy effective string theory there is a scalar field $\phi$ called the dilaton (that is coupled to the Ricci scalar R) of form $F(\phi)R$ \cite{gas1993}. Furthermore, the higher-order string correction contains a non-canonical kinetic term of form $(\nabla\phi)^4$ as well as a field self-interaction of form $\xi(\phi)(\nabla\phi)^2\Box\phi$ in the action. We accommodate non-linear field derivative terms using a Lagrangian of form $K(\phi,X)-G(\phi,X)\Box\phi$ \cite{def2010}.

We compare observed data of the growth rate of matter density perturbations to theoretical predictions. Whereas the background expansion history in modified gravity is nearly identical to that of dark energy models, the evolution of matter density perturbations in modified gravity is different from that of dark energy models (the cosmological constant is the standard candidate for dark energy). Thus, it is important to study the growth history of perturbations to distinguish modified gravity from models based on the cosmological constant or dark energy. Therefore, we computed the growth rate of matter density perturbations in Galileon cosmology and compared it to observational data. In addition, we studied the cosmological constraints on the Galileon model obtained from SN Ia, CMB anisotropies, and BAO observations so as to test the validity of the model.

This paper is organized as follows. In the next section, we describe the background evolution and the evolution of linear perturbations in Galileon cosmology. In Sec. \ref{constraint} we study the cosmological constraints on Galileon gravity obtained from observational data of the growth rate of matter density perturbations, SN Ia, CMB, and BAO. Finally, conclusions are given in Sec. \ref{conclusion}.

\section{MODEL \label{model}}
\subsection{Background evolution}
The action we consider is of the form
%\begin{widetext}
\begin{equation}
S=\int{d^4x\sqrt{-g}\left[\frac{M_{\rm pl}^2}{2}F(\phi)R+K(\phi,X)-G(\phi,X)\Box\phi+L_m\right]}, \label{action}
\end{equation}
%\end{widetext}
where $g$ is the determinant of the space-time metric $g_{\mu\nu}$, $M_{\rm pl}=(8\pi G)^{-1/2}$ is the reduced Planck mass, $G$ is the gravitational constant, $R$ is the Ricci scalar, and $\phi$ is a scalar field with a kinetic term $X=-(1/2)(\nabla\phi)^2$. The function $F(\phi)$ depends on $\phi$ only, whereas $K(\phi,X)$ and $G(\phi,X)$ are functions of both $\phi$ and $X$. $L_m$ is the matter Lagrangian. We have used the notation: $(\nabla\phi)^2=g^{\mu\nu}\nabla_\mu\phi\nabla_\nu\phi$, $\Box\phi=g^{\mu\nu}\nabla_\mu\nabla_\nu\phi$.

Variation with respect to the metric produces the Einstein equations; variation with respect to the Galileon field $\phi$ yields the equation of motion. For Friedmann--Robertson--Walker spacetime, the Einstein equations give
\vspace{0.1mm}
%\begin{widetext}
\begin{equation}
3M_{\rm pl}^2FH^2=\rho_m+\rho_r-3M_{\rm pl}^2H\dot{F}-K+2XK_{,X}+6H\dot{\phi}XG_{,X}-2XG_{,\phi}, \label{freq1}
\end{equation}
\begin{equation}
-M_{\rm pl}^2F(3H^2+2\dot{H})=p_r+2M_{\rm pl}^2H\dot{F}+M_{\rm pl}^2\ddot{F}+K-2XG_{,X}\ddot{\phi}-2XG_{,\phi}, \label{freq2}
\end{equation}
and the equation of motion for the Galileon field gives
\begin{eqnarray}
& & (K_{,X}+2XK_{,XX}+6H\dot{\phi}G_{,X}+6H\dot{\phi}XG_{,XX}-2XG_{,\phi X}-2G_{,\phi})\ddot{\phi} \nonumber \\
& & +(3HK_{,X}+\dot{\phi}K_{,\phi X}+9H^2\dot{\phi}G_{,X}+3\dot{H}\dot{\phi}G_{,X}+6HXG_{,\phi X}-6HG_{,\phi}-G_{,\phi\phi}\dot{\phi})\dot{\phi} \nonumber \\
& & -K_{,\phi}-6M_{\rm pl}^2H^2F_{,\phi}-3M_{\rm pl}^2\dot{H}F_{,\phi}=0,
\end{eqnarray}
%\end{widetext}
where an overdot represents differentiation with respect to cosmic time $t$ and $H=\dot{a}/a$ is the Hubble expansion rate. Note that we have used the following partial derivative notation: $K_{,X}\equiv\partial K/\partial X$ and $K_{,XX}\equiv\partial^2K/\partial X^2$ (and similarly for other variables).
$\rho_m$ and $\rho_r$ are the energy densities of matter and radiation, respectively, and $p_r$ is the pressure of the radiation.

In this paper, as we are interested in the basic parameters of Galileon theory, we consider the following functions
\begin{equation}
F(\phi)=\frac{2}{M_{\rm pl}^2}\phi, \label{scf1}
\end{equation}
\begin{equation}
K(\phi,X)=2\frac{\omega}{\phi}X, \label{scf2}
\end{equation}
\begin{equation}
G(\phi,X)=2\xi(\phi)X, \label{scf3}
\end{equation}
\newpage
where $\omega$ is the Brans--Dicke parameter and $\xi(\phi)$ is a function of $\phi$. This model is the Brans--Dicke theory with the following self-interaction term: $\xi(\phi)(\nabla\phi)^2\Box\phi$.

In this case, the Friedmann equation (\ref{freq1}) and (\ref{freq2}) can be written in the following forms, respectively,
\begin{equation}
3H^2=\frac{1}{M_{\rm pl}^2}(\rho_m+\rho_r+\rho_\phi) \label{frefre1},
\end{equation}
\begin{equation}
-3H^2-2\dot{H}=\frac{1}{M_{\rm pl}^2}(p_r+p_\phi) \label{frefre2},
\end{equation}
where the effective dark energy density $\rho_\phi$ is defined as
\begin{equation}
\rho_\phi = 2\phi\left[-3H\frac{\dot{\phi}}{\phi}+\frac{\omega}{2}\left(\frac{\dot{\phi}}{\phi}\right)^2+\phi^2\xi(\phi)\left\{3H+\frac{\dot{\phi}}{\phi}\right\}\left(\frac{\dot{\phi}}{\phi}\right)^3\right]+3H^2\left(M_{\rm pl}^2-2\phi\right), \label{rho_phi}
\end{equation}
and the effective pressure of dark energy $p_\phi$ is
\begin{equation}
p_\phi = 2\phi\left[\frac{\ddot{\phi}}{\phi}+2H\frac{\dot{\phi}}{\phi}+\frac{\omega}{2}\left(\frac{\dot{\phi}}{\phi}\right)^2-\phi^2\xi(\phi)\left\{\frac{\ddot{\phi}}{\phi}-\left(\frac{\dot{\phi}}{\phi}\right)^2\right\}\left(\frac{\dot{\phi}}{\phi}\right)^2\right]-(3H^2+2\dot{H})\left(M_{\rm pl}^2-2\phi\right). \label{p_phi}
\end{equation}

For the numerical analysis in this paper, we adopt a specific model in which
\begin{equation}
\xi(\phi)=\frac{r_c^2}{\phi^2}, \label{scf4}
\end{equation}
where $r_c$ is the crossover scale \cite{cho2009}. This coincides with the model by Refs. \citen{sil2009,kob2010a}. Evolution of matter density perturbations for this model has been computed in Refs. \citen{sil2009,kob2010a}.

At early times, to recover general relativity, we set the initial condition $\phi\simeq M_{\rm pl}^2/2$. This reduces the Einstein equations (\ref{frefre1}) and (\ref{frefre2}) to their usual forms: $3H^2 = {(\rho_m+\rho_r)}/M_{\rm pl}^2$ and $-3H^2 - 2\dot{H}={p_r}/M_{\rm pl}^2$. This is the cosmological version of the Vainshtein effect, the method by which general relativity is recovered below a certain scale. At present, to describe the cosmic acceleration, the value of $r_c$ must be fine-tuned.

\begin{figure}
\begin{center}
\includegraphics[width=120mm]{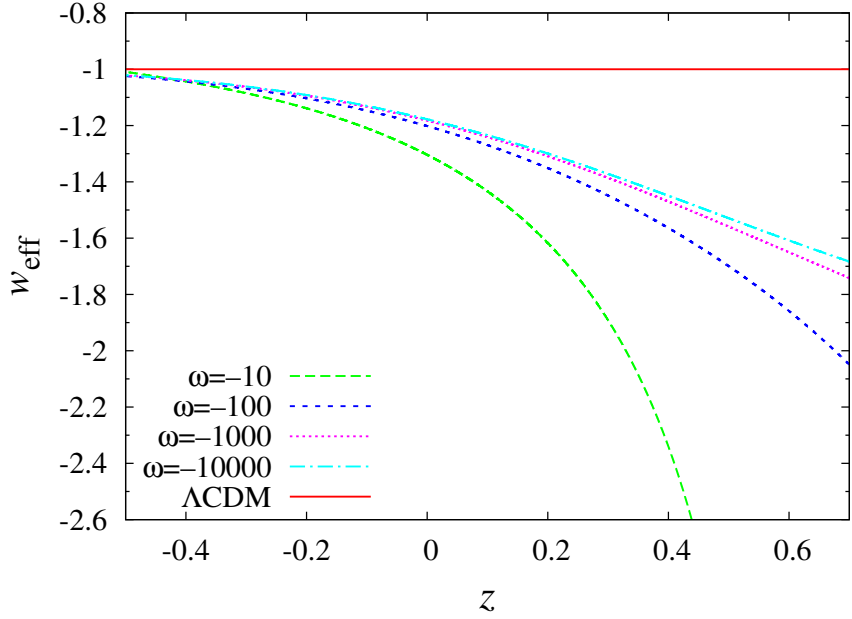}
\caption{Effective equation of state $w_{\rm eff}$ in Galileon gravity as a function of redshift $z$ for various values of the Brans--Dicke parameter $\omega$. The parameters are given by $\Omega_{m,0}=0.30$. \label{weff}}
\end{center}
\end{figure}

In Fig. \ref{weff}, we plot the effective equation of state $w_{\rm eff}$ in Galileon gravity, which is described as $w_{\rm eff}=-\dot{\rho_\phi}/(3H\rho_\phi)-1$ using the effective dark energy density $\rho_\phi$ in Eq. (\ref{rho_phi}). We set $\Omega_{m,0}=0.30$ for the Galileon models in Fig. \ref{weff}. We can see the asymptotic behavior $w_{\rm eff} \rightarrow -1$ for the future. A phantom-like behavior is found at low $z$. The smaller absolute value of the Brans--Dicke parameter $|\omega |$ ($\omega < 0$) is, the smaller $w_{\rm eff}$ is at low $z$. The model in our paper is the Brans-Dicke theory extended by adding the following self-interaction term: $\xi(\phi)(\nabla\phi)^2\Box\phi$. Thus, $\omega$ of this model is not exactly the same as the original Brans-Dicke parameter. If $|\omega|$ is infinity with fixing the value of $\Omega_{m,0}$, crossover scale $r_c$ also approaches infinity, and the terms including $\xi(\phi)$ in Eqs. (\ref{rho_phi}) and (\ref{p_phi}) behave non-trivially. In the case of fixing the value of $\Omega_{m,0}$, even if $|\omega|$ is infinity, this model does not approach $\Lambda$CDM.

\subsection{Density perturbations}
The evolution equation for the cold dark matter overdensity $\delta$ in linear theory is governed by
\begin{equation}
\ddot{\delta}+2H\dot{\delta}-4\pi G_{\rm eff}\rho\delta\simeq0,
\end{equation}
where $G_{\rm eff}$ represents the effective Newton's constant in Galileon gravity. For the model specified by Eqs. (\ref{scf1}), (\ref{scf2}), and (\ref{scf3}), the effective Newton's constant is given by
\begin{equation}
G_{\rm eff}=\frac{1}{16\pi\phi}\left[1+\frac{(1+\xi(\phi)\dot{\phi}^2)^2}{J}\right],
\end{equation}
where
\begin{equation}
J=3+2\omega+\phi^2\xi(\phi)\left[4\frac{\ddot{\phi}}{\phi}-2\frac{\dot{\phi}^2}{\phi^2}+8H\frac{\dot{\phi}}{\phi}-\phi^2\xi(\phi)\frac{\dot{\phi}^4}{\phi^4}\right].
\end{equation}
The effective Newton's constant $G_{\rm eff}$ is close to Newton's constant $G$ at early times, but increases at later times.

\begin{figure}
\begin{center}
\includegraphics[width=120mm]{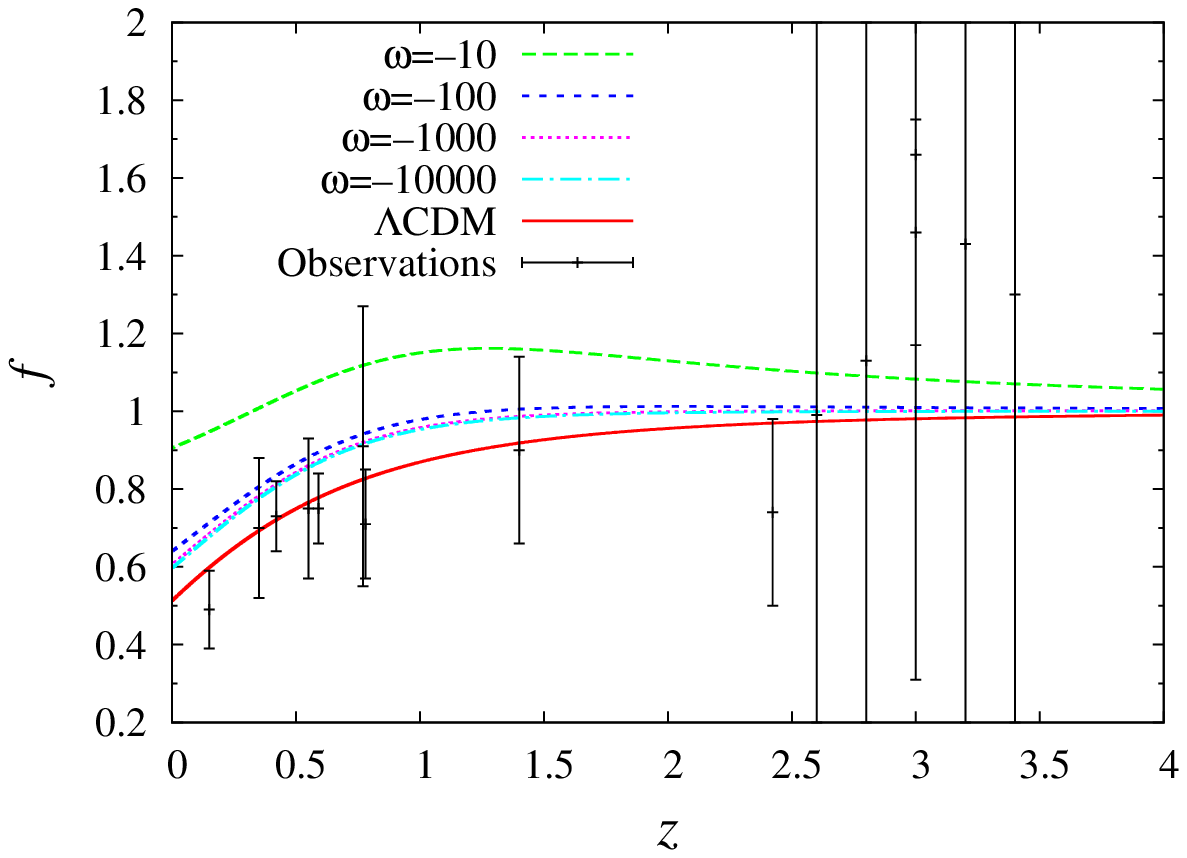}
\caption{Growth rate $f$ in Galileon gravity as a function of redshift $z$ for various values of the Brans--Dicke parameter $\omega$. The parameters are given by $\Omega_{m,0}=0.30$. \label{f_w}}
%\end{figure}
%\begin{figure}[h!]
\includegraphics[width=120mm]{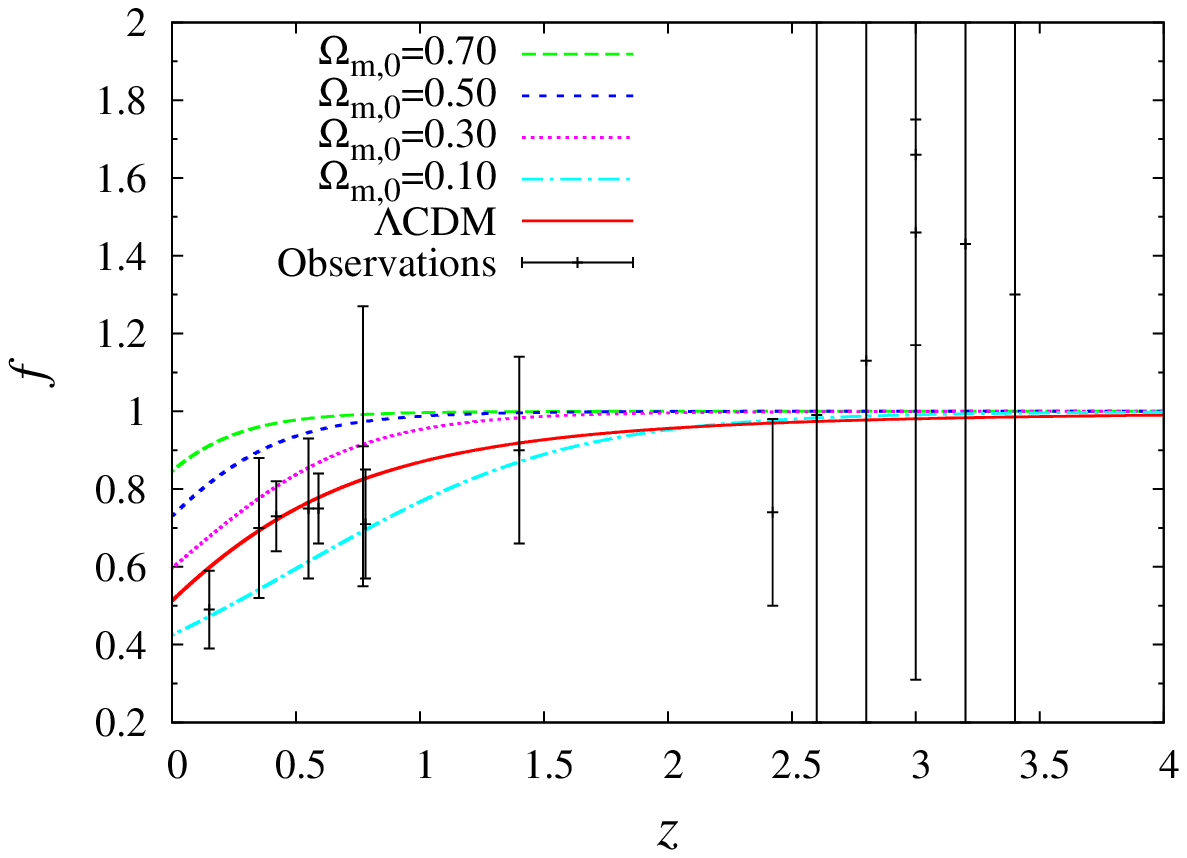}
\caption{Growth rate $f$ in Galileon gravity as a function of redshift $z$ for various values of today's energy density parameter of matter $\Omega_{m,0}$. The parameters are given by $\omega=-10000$. \label{f_m}}
\end{center}
\end{figure}

We set the initial conditions $\delta\approx a$, $\dot{\delta}\approx\dot{a}$ at early times. As we are interested in the difference between the growth of density perturbations in the Galileon gravity and that in the $\Lambda$CDM model, we assume that initial conditions of matter density perturbations are same as the conventional $\Lambda$CDM case. Solving the evolution equation numerically, we obtain the growth factor $\delta/a$ for Galileon gravity.
The linear growth rate is written as
\begin{equation}
f=\frac{{\rm d}\ln{\delta}}{{\rm d}\ln{a}}.
\end{equation}
The growth rate can be parameterized by the growth index $\gamma$, as defined by
\begin{equation}
f=\Omega_m^\gamma.
\end{equation}
The matter energy density parameter is defined as $\Omega_{m}=\rho_m/3M_{\rm pl}^2H^2$.

We plot the growth rate $f$ in Galileon gravity in Fig. \ref{f_w} and Fig. \ref{f_m}. $\omega$ is the (constant) Brans--Dicke parameter and $\Omega_{m,0}$ is the energy density parameter of matter at the present day. We set $\Omega_{m,0}=0.30$ for the Galileon model in Fig. \ref{f_w} and $\omega=-10000$ for the Galileon model in Fig. \ref{f_m}. $\Omega_{m,0}$ for the $\Lambda$CDM model is $0.30$ in these figures.

For the same value of $\Omega_{m,0}$, the growth rate $f$ in a Galileon model is enhanced compared to the $\Lambda$CDM case, because of the enhancement of Newton's constant. The smaller $\Omega_{m,0}$ is, the more suppressed the growth rate is. We describe the observational data of Fig. \ref{f_w} and Fig. \ref{f_m} in the next section.

\section{OBSERVATIONAL CONSTRAINTS \label{constraint}}

In this section, we study the cosmological constraints on Galileon gravity so as to test the validity of the model.
We carried out a detailed investigation of the allowed parameter region using the following observational data.

\subsection{Observational data}

\subsubsection{Growth rate}

In Table \ref{fdata} we list the observational data used for the growth factor $f$ from galaxy redshift distortions and Lyman--$\alpha$ forests, as compiled in Refs. \citen{por2008,nes2008,guz2008}, adding the data from The WiggleZ Dark Energy Survey \cite{bla2010} from original references, provided to the right of the table.

\begin{table}
\begin{center}
\caption{Currently available data for linear growth rates $f_{obs}$ used in our analysis. $z$ is redshift; $\sigma$ is the 1$\sigma$ uncertainty of the growth rate data. \label{fdata}}
%\begin{ruledtabular}
\begin{tabular}{c c c c c}
\hline
$z$ & $f_{obs}$ & $\sigma$ & Ref. \\
\hline\hline
0.15 & 0.49~ & 0.1 & \citen{guz2008,col2001} \\
0.35 & 0.7~ & 0.18 & \citen{teg2006} \\
0.42 & 0.73~ & 0.09 & \citen{bla2010} \\
0.55 & 0.75~ & 0.18 & \citen{ros2007} \\
0.59 & 0.75~ & 0.09 & \citen{bla2010} \\
0.77 & 0.91~ & 0.36 & \citen{guz2008} \\
0.78 & 0.71~ & 0.14 & \citen{bla2010} \\
1.4 & 0.9~ & 0.24 & \citen{ang2008} \\
3.0 & 1.46~ & 0.29 & \citen{mac2005} \\
2.125 -- 2.72 & 0.74~ & 0.24 & \citen{viel2004} \\
2.2 -- 3 & 0.99~ & 1.16 & \citen{viel2006} \\
2.4 -- 3.2 & 1.13~ & 1.07 & \citen{viel2006} \\
2.6 -- 3.4  & 1.66~ & 1.35 & \citen{viel2006} \\
2.8 -- 3.6 & 1.43~ & 1.34 & \citen{viel2006} \\
3 -- 3.8 & 1.3~ & 1.5 & \citen{viel2006} \\
\hline
\end{tabular}
%\end{ruledtabular}
\end{center}
\end{table}

The statistical $\chi^2$ function for the growth rate is defined as
\begin{equation}
\chi^2_{f}=\sum_{i=1}^{15}\frac{(f(z_i)-f_{obs}(z_i))^2}{\sigma(z_i)^2}, \label{chi2}
\end{equation}
where $\sigma(z)$ is the 1$\sigma$ uncertainty in the $f_{obs}(z)$ data.

\subsubsection{Type Ia supernovae}

The Union2 SN Ia data \cite{ama2010} consists of low $z$ SN Ia data observed at the F.L. Whipple observatory of the Harvard--Smithsonian center for astrophysics \cite{hic2009}, intermediate $z$ data observed during the first season of the Sloan Digital Sky Survey (SDSS)-II supernova survey \cite{kessler2010}, and high $z$ data from the Union compilation \cite{kow2008}. The Union2 SN Ia data used the SALT2 light curve fitter because it performs better than either SALT or MLCS2k2 in terms of the scatter around the best-fit luminosity distance relation \cite{ama2010}.

The 557 Union2 SN Ia data set \cite{ama2010} gives the distance modulus at redshift $\mu_{obs}(z_i)$.

The theoretical distance modulus (for a given model) is defined by
\begin{equation}
\mu(z)=5\log_{10}{D_L}+\mu_0,
\end{equation}
where $D_L$ is the Hubble-free luminosity distance given by
\begin{equation}
D_L=(1+z)\int^z_0\frac{H_0}{H(z^{\prime})}dz^{\prime},
\end{equation}
and $\mu_0$ is
\begin{equation}
\mu_0=5\log_{10}{\left(\frac{{H_0}^{-1}}{Mpc}\right)} + 25 = 42.38 - 5\log_{10}{h}, \label{eq:mu}
\end{equation}
where $h$ is the Hubble constant $H_0$ in units of $100~{\rm km~s^{-1}~Mpc^{-1}}$.

For the SN Ia data we have
\begin{equation}
\chi_{\rm SN Ia}^2=\sum^{557}_{i=1}\left[\frac{\mu(z_i) - \mu_{obs}(z_i)}{\sigma_{\mu}(z_i)}\right]^2,
\end{equation}
where $\sigma_{\mu}$ is the total uncertainty in the distance modulus.

As the nuisance parameter $\mu_0$ (Eq. (\ref{eq:mu})) is model-independent, we analytically marginalize it as follows:
\begin{equation}
\chi_{\rm SN Ia}^2=a-\frac{b^2}{c},
\end{equation}
where
\begin{equation}
a=\sum^{557}_{i=1}\frac{[\mu(z_i) - \mu_{obs}(z_i)]^2}{\sigma_{\mu}^2(z_i)},
\end{equation}
\begin{equation}
b=\sum^{557}_{i=1}\frac{\mu(z_i) - \mu_{obs}(z_i)}{\sigma_{\mu}^2(z_i)},
\end{equation}
and
\begin{equation}
c=\sum^{557}_{i=1}\frac{1}{\sigma_{\mu}^2(z_i)}.
\end{equation}

\subsubsection{Cosmic microwave background}

The positions of CMB acoustic peaks are affected by the expansion history of the Universe from the decoupling epoch up to today. To quantify the shift of acoustic peaks we use data points $l_a$, $R$, $z_*$ of Ref. \citen{kom2010} (WMAP7), where $l_a$ and $R$ are two CMB shift parameters and $z_*$ is the redshift at decoupling. For the flat Universe, we have
\begin{equation}
R = \sqrt{\Omega_{m,0}}H_0d_a(z_*),
\end{equation}
where $d_a(z_*)=\int^{z_*}_0{dz/H(z)}$ is the comoving angular diameter distance.

The multipole $l_a$ is defined by
\begin{equation}
l_a = \pi\frac{d_a(z_*)}{r_s(z_*)},
\end{equation}
where $r_s(z_*)=\int^{\infty}_{z_*}dz/[H(z)\sqrt{3\{1+3\Omega_{b,0}/(4\Omega_{\gamma,0}(1+z))\}}]$ is the sound horizon at the decoupling. $\Omega_{b,0}$ and $\Omega_{\gamma,0}$ are today's density parameters of baryons and photons, respectively.

For the redshift $z_*$ there is a fitting formula by Hu and Sugiyama \cite{hus1996}:
\begin{equation}
z_*=1048[1+0.00124(\Omega_{b,0}{h^2})^{-0.738}][1+g_1(\Omega_{m,0}{h^2})^{g_2}],
\end{equation}
where
\begin{equation}
g_1=\frac{0.0783(\Omega_{b,0}{h^2})^{-0.238}}{1+39.5(\Omega_{b,0}{h^2})^{0.763}},
\end{equation}
\begin{equation}
g_2=\frac{0.560}{1+21.1(\Omega_{b,0}{h^2})^{1.81}}.
\end{equation}
In our analysis of the Galileon model, we adopt the values $h=0.71$ and $\Omega_{b,0} = 0.02258 h^{-2}$ \cite{kom2010}, and the standard value $\Omega_{\gamma,0} = 2.469 \times 10^{-5} h^{-2}$.

For a flat prior, the 7-year WMAP data measured best-fit values are \cite{kom2010}
\begin{equation}
{\textnormal{\mathversion{bold}$\bar{V}$}}_{\rm CMB}
=\left( \begin{array}{c}
        \bar{l_a} \\
        \bar{R} \\
        z_*
        \end{array} \right)
=\left( \begin{array}{c}
        302.09 \pm 0.76 \\
        1.725 \pm 0.018 \\
        1091.3 \pm 0.91
        \end{array} \right).
\end{equation}
The corresponding inverse covariance matrix is \cite{kom2010}
\begin{equation}
{\textnormal{\mathversion{bold}$C$}}^{-1}_{\rm CMB}
=\left( \begin{array}{ccc}
        2.305  &  29.698  &  -1.333 \\
        29.698  &  6825.270  &  -113.180 \\
        -1.333  &  -113.180  &  3.414
        \end{array} \right).
\end{equation}
Thus, we define
\begin{equation}
{\textnormal{\mathversion{bold}$X$}}_{\rm CMB}
=\left( \begin{array}{c}
        l_a - 302.09 \\
        R - 1.725 \\
        z_* - 1091.3
        \end{array} \right),
\end{equation}
and find the contribution of CMB is
\begin{equation}
\chi^2_{\rm CMB}={\textnormal{\mathversion{bold}$X$}}^T_{\rm CMB}{\textnormal{\mathversion{bold}$C$}}^{-1}_{\rm CMB}{\textnormal{\mathversion{bold}$X$}}_{\rm CMB}.
\end{equation}

\subsubsection{Baryon acoustic oscillation}

For BAO analysis, we apply the maximum likelihood method using the data points of Ref. \citen{per2010} (SDSS7):
\begin{equation}
{\textnormal{\mathversion{bold}$\bar{V}$}}_{\rm BAO}
=\left( \begin{array}{c}
        \frac{r_s(z_d)}{D_V(0.2)} = 0.1905 \pm 0.0061 \\
        \frac{r_s(z_d)}{D_V(0.35)} = 0.1097 \pm 0.0036
        \end{array} \right),
\end{equation}
where $r_s(z_d)$ is the sound horizon at the baryon drag epoch $z_d$. For $z_d$, we use the fitting formula by Eisenstein and Hu \cite{Eisenstein1998}:
\begin{equation}
z_d = \frac{1291(\Omega_{m,0}h^2)^{0.251}}{1+0.659(\Omega_{m,0}h^2)^{0.828}}[1+b_1(\Omega_{b,0}h^2)^{b_2}],
\end{equation}
where
\begin{equation}
b_1 = 0.313(\Omega_{m,0}h^2)^{-0.419}[1+0.607(\Omega_{m,0}h^2)^{0.674}],
\end{equation}
and
\begin{equation}
b_2 = 0.238(\Omega_{m,0}h^2)^{0.223}.
\end{equation}
The dilation scale $D_V$ at the redshift z is
\begin{equation}
D_V(z) = \left[{d_a(z)}^2\frac{z}{H(z)}\right]^{1/3},
\end{equation}
where $d_a(z)$ is the comoving angular diameter distance.

Thus, we construct
\begin{equation}
{\textnormal{\mathversion{bold}$X$}}_{\rm BAO}
=\left( \begin{array}{c}
        \frac{r_s(z_d)}{D_V(0.2)} - 0.1905 \\
        \frac{r_s(z_d)}{D_V(0.35)} - 0.1097
        \end{array} \right),
\end{equation}
and using the inverse covariance matrix \cite{per2010},
\begin{equation}
{\textnormal{\mathversion{bold}$C$}}^{-1}_{\rm BAO}
=\left( \begin{array}{cc}
        30124  &  −17227 \\
        −17227  &  86977
        \end{array} \right).
\end{equation}
The contribution of BAO to $\chi^2$ is
\begin{equation}
\chi^2_{\rm BAO}={\textnormal{\mathversion{bold}$X$}}^T_{\rm BAO}{\textnormal{\mathversion{bold}$C$}}^{-1}_{\rm BAO}{\textnormal{\mathversion{bold}$X$}}_{\rm BAO}.
\end{equation}

To determine the best value for, and the allowed region of, the parameters, we use the maximum likelihood method.

\subsection{Numerical results}
We now present our main results for the observational constraints on the Galileon gravity specified by Eqs. (\ref{scf1}), (\ref{scf2}), (\ref{scf3}), and (\ref{scf4}).

\begin{figure}
\begin{center}
\includegraphics[width=120mm]{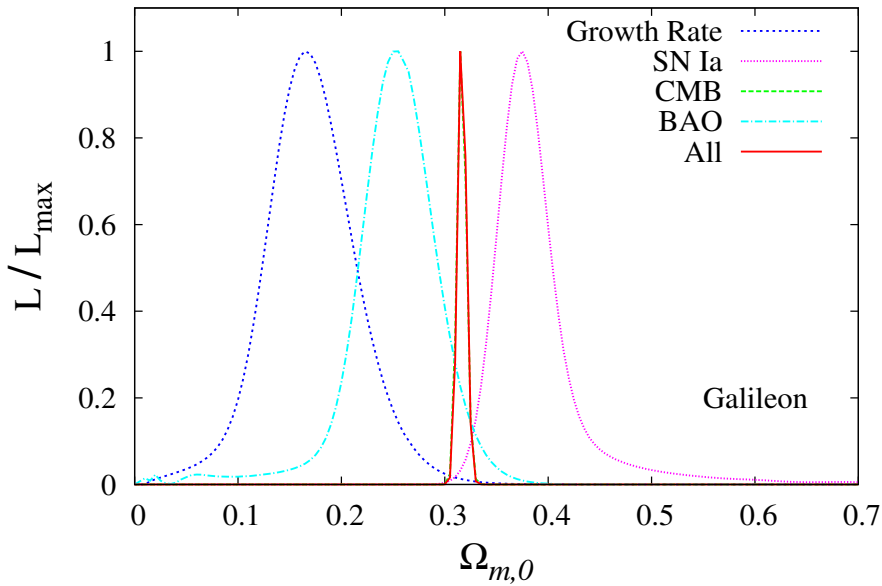}
\caption{1D probability distribution of the energy density parameter of matter $\Omega_{m,0}$ for the Galileon model from observational data of Growth Rate, SN Ia, CMB, BAO, respectively, and the combination of All data. The line of CMB almost overlaps with the line of All. \label{1domegam}}
\end{center}
\end{figure}

In Fig. \ref{1domegam}, we plot the probability distribution of the energy density parameter of matter $\Omega_{m,0}$ in the Galileon model from observational data of (individually) growth rate, SN Ia, CMB, and BAO, and also the combination of all data, where the other parameter is marginalized.

Considering only the growth rate data, the best fit value for the Galileon model is $\Omega_{m,0}$ = 0.166. This is considerably smaller than value obtained from observations of SN Ia, CMB, and BAO. Using only the growth rate data, we obtained the following constraint:
\begin{equation}
\Omega_{m,0}=0.166^{+0.044}_{-0.039}.~~~~~~~~(68.3\%~{\rm C.L.})~~~~~~~~{(\rm growth~rate~data)}. \label{om_growth}
\end{equation}

By contrast, using the combination of growth rate, SN Ia, CMB, and BAO data, the best fit value for the Galileon model is $\Omega_{m,0}$ = 0.317; we obtained the stringent constraint as follows:
\begin{equation}
\Omega_{m,0}=0.317^{+0.003}_{-0.004}.~~~~~~~~(68.3\%~{\rm C.L.})~~~~~~~~{(\rm all~data)}. \label{om_all}
\end{equation}
In the Galileon model, the bounds produced by each set of observations are inconsistent.
 
\begin{figure}
\begin{center}
\includegraphics[width=120mm]{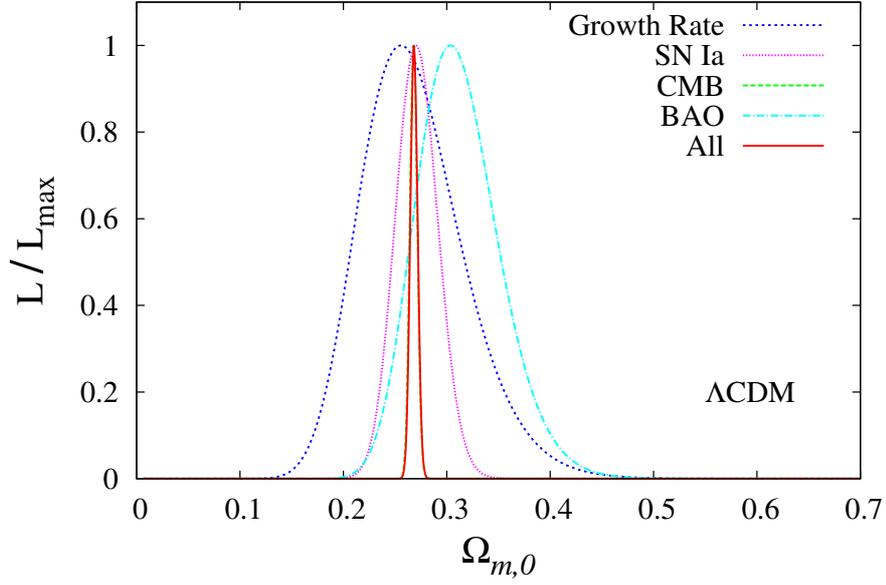}
\caption{1D probability distribution of $\Omega_{m,0}$ for the $\Lambda$CDM model from observational data of growth rate, SN Ia, CMB, and BAO, individually, and the combination of all data. The line of CMB almost overlaps with the line of All. \label{1dlcdm}}
\end{center}
\end{figure}

 In Fig. \ref{1dlcdm}, we plot the probability distribution of $\Omega_{m,0}$ for the $\Lambda$CDM model from observational data of (individually) growth rate, SN Ia, CMB, and BAO, and also the combination of all data.

In the $\Lambda$CDM model, the bounds from each observational data set are consistent.

\begin{figure}
\begin{center}
\includegraphics[width=120mm]{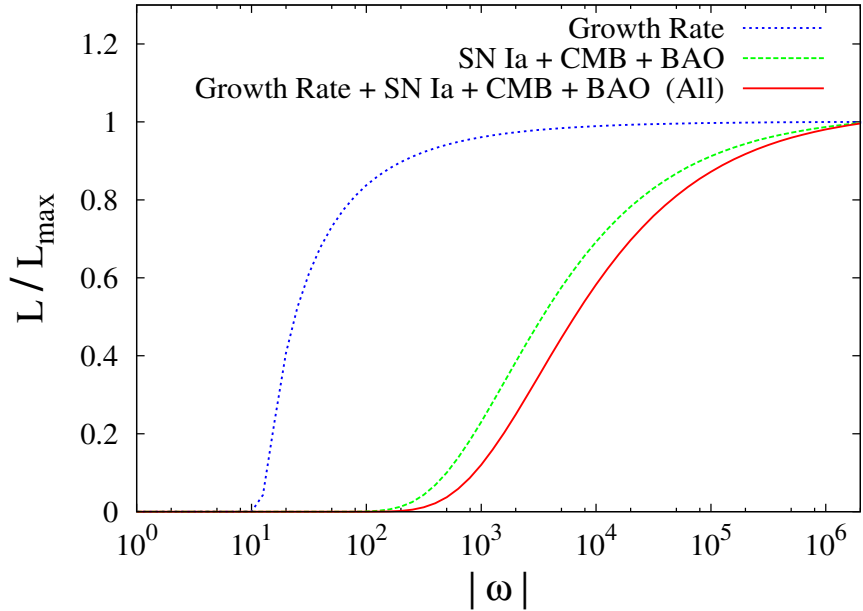}
\caption{1D probability distribution of absolute value of the Brans--Dicke parameter $|\omega |$ (note that $\omega < 0$) for the Galileon model from observational data of growth rate, the combination of SN Ia, CMB, BAO data, and the combination of all data. \label{1dxi}}
\end{center}
\end{figure}

In Fig. \ref{1dxi}, we plot the probability distribution of the Brans--Dicke parameter $\omega$ for the Galileon model from observational data of growth rate, the combination of SN Ia, CMB, BAO data, and the combination of all data, where other parameter is marginalized.

Using only the growth rate data, we obtained the following constraint:
\begin{equation}
\omega < -1060.~~~~~~~~(68.3\%~{\rm C.L.})~~~~~~~~{(\rm growth~rate~data)}.
\end{equation}
Furthermore, from the combination of growth rate, SN Ia, CMB, and BAO data (all data), we obtained the stringent constraint as follows:
\begin{equation}
\omega < -36900.~~~~~~~~(68.3\%~{\rm C.L.})~~~~~~~~{(\rm all~data)}.
\end{equation}

\begin{figure}
\begin{center}
\includegraphics[width=120mm]{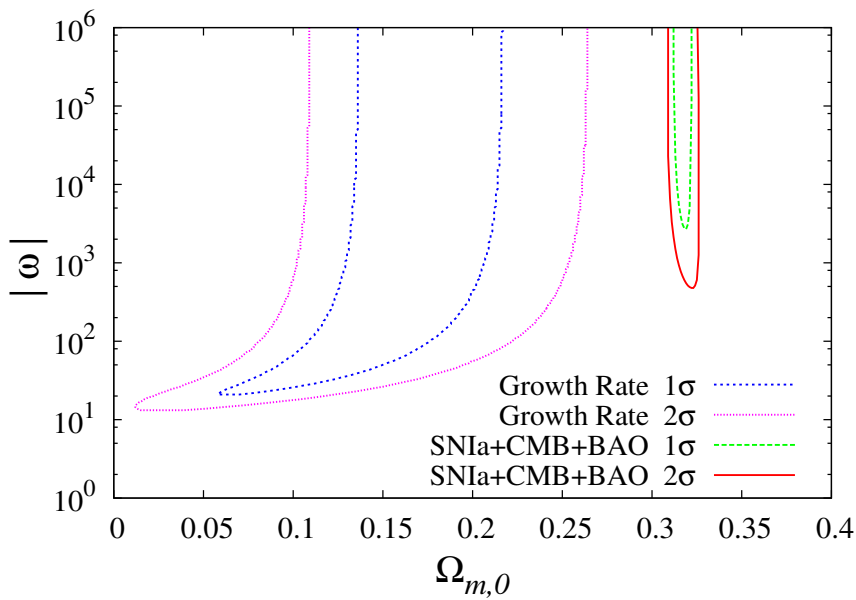}
\caption{Probability contours in the ($\Omega_{m,0}$, $|\omega |$)-plane for the Galileon model. The contours show the 1$\sigma$ (68.3\%) and 2$\sigma$ (95.0\%) confidence limits, from the growth rate data, and from the combination of SN Ia, CMB, and BAO data, respectively. \label{2d}}
\end{center}
\end{figure}

In Fig. \ref{2d}, we plot the probability contours in the ($\Omega_{m,0}$, $|\omega |$)-plane in the Galileon model.
The contours show the 1$\sigma$ (68.3\%) and 2$\sigma$ (95.0\%) confidence limits, from the growth rate data, and from the combination of SN Ia, CMB, and BAO data. ($|\omega |$ is the absolute value of the Brans--Dicke parameter; $\omega < 0$.)

In the Galileon model, the allowed parameter region obtained using only the growth rate data does not overlap with the allowed parameter region obtained from the combination of SN Ia, CMB, and BAO data at all.

In Table \ref{table2}, we list the best fit parameters, the $\chi^2$ values, and the differences of the Akaike information criteria (AIC) \cite{aka1974} and the Bayesian information criteria (BIC) \cite{sch1978}
for both the Galileon model and the $\Lambda$CDM model. Here, we use the combination of growth rate, SN Ia, CMB, and BAO data (all data).
The $\chi^2$ value, AIC change, and BIC change for the Galileon model are larger than those of the $\Lambda$CDM model. In the Galileon model, as the bounds from growth rate data is inconsistent with the bounds from other observational data, the $\chi^2$ value for all data is large. In the analysis of this paper, we find that the Galileon model is less compatible with observations than is the $\Lambda$CDM model. This result seems to be qualitatively the same in most of the generalized Galileon models in which Newton's constant is enhanced.

\begin{table}
\begin{center}
\caption{Results of observational tests using growth rate, SN Ia, CMB, and BAO data (all data). We consider the spatially flat Universe. \label{table2}}
\begin{tabular}{l l c r r}
\hline
Model  &  Best fit parameters  &  $\chi^2$  &  $\Delta$AIC  &  $\Delta$BIC \\
\hline\hline
$\Lambda$CDM  &  $\Omega_{m,0}=0.268$ ~ &  550.674 ~ & 0.00  &  0.00 \\
\hline
Galileon ~ &  $\Omega_{m,0}=0.316$ ~ &  582.013 ~ & 33.339 & ~ 37.697   \\
 &  $\omega=-2.69\times 10^6$ & & & \\
\hline
\end{tabular}
\end{center}
\end{table}

\begin{figure}
\begin{center}
\includegraphics[width=120mm]{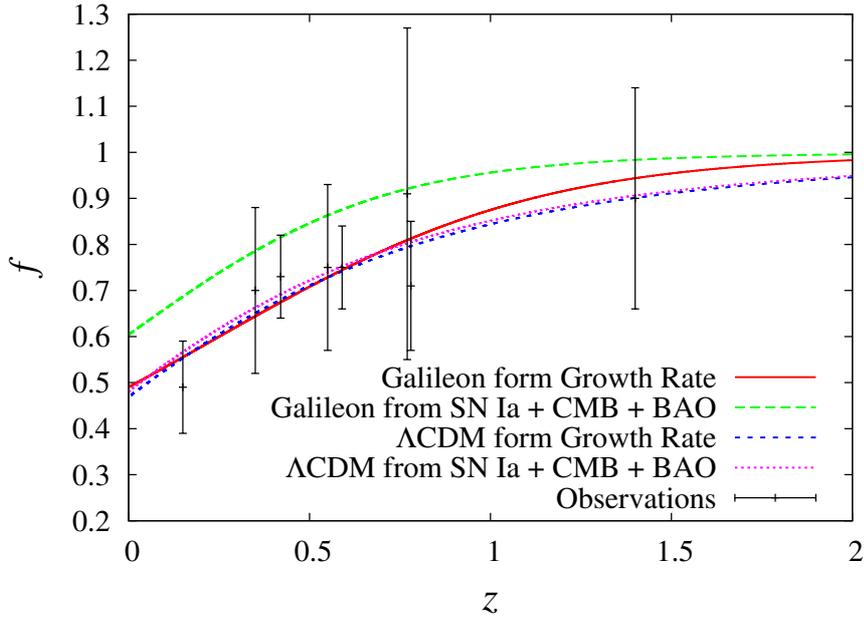}
\caption{Growth rate $f$ in best fit models of Galileon theory and $\Lambda$CDM from only the growth rate data, and from the combination of SN Ia, CMB, and BAO data. \label{best}}
\end{center}
\end{figure}

In Fig. \ref{best}, we plot the growth rate $f$ using the best fit models of the Galileon theory and $\Lambda$CDM using only the growth rate data, and using the combination of SN Ia, CMB, and BAO data.

For the Galileon model obtained from the combination of SN Ia, CMB, and BAO data (excluding growth rate data), the growth rate of matter density perturbations is enhanced. On the other hand, in the $\Lambda$CDM case, The model using only growth rate data is consistent with the model using the combination of SN Ia, CMB, and BAO data.

\section{CONCLUSIONS \label{conclusion}}

For the same value of $\Omega_{m,0}$, the growth rate $f$ in Galileon models is enhanced compared to that of the $\Lambda$CDM case, because of the enhancement of Newton's constant. The smaller $\Omega_{m,0}$ is, the more suppressed growth rate is. Therefore, the best fit value of $\Omega_{m,0}$ in the Galileon model using only the growth rate data is $\Omega_{m,0}$ = 0.166, a value that is relatively small.

By contrast, the value of $\Omega_{m,0}$ in the Galileon model obtained from the combination of SN Ia, CMB, and BAO data is larger than that in the $\Lambda$CDM case. This result is consistent with Refs. \citen{nes2010,kim2010}. In Refs. \citen{nes2010,kim2010}, the cosmological constraints using observational data of SN Ia, CMB, and BAO have been studied, but there the models are not the same as the one in our analysis. In this paper, we add the important constraints obtained from the growth rate data of matter density perturbations, compare the allowed parameter region from the growth rate data to the region from the combination of SN Ia, CMB, and BAO data.

As seen in Fig. \ref{2d}, in the Galileon model, the bound based only on the growth rate data is completely incompatible with the bound based on the combination of SN Ia, CMB, and BAO data. This result seems to be qualitatively the same in most of the generalized Galileon models in which Newton's constant is enhanced. On the other hand, in the $\Lambda$CDM model, the bounds based on each observation data set are consistent.

We also find that the upper limit of the Brans--Dicke parameter $\omega$ for the Galileon model is not constrained tightly only by growth rate data, but that the stringent upper limit of $\omega$ is obtained from the combination of growth rate, SN Ia, CMB, and BAO data.

Galileon gravity is a fascinating model that can induce the accelerated expansion of the present Universe, meets the requirements of solar system experiments, and can avoid ghost and instability problems. However, through the analysis presented in this paper, we found that the Galileon model is less compatible with observations than the $\Lambda$CDM model is.

As the constraint from (only) growth rate data is not tight, more and better growth rate data are required to distinguish between dark energy and modified gravity.
%\bibliography{hirano}

%\section*{Acknowledgements}
%We would like to thank ...........

%\appendix
%\section{First Appendix} %Empty argument \section{} yields `Appendix'. 
%
%\section{Second Appendix}

\end{document}